## TITLE

Adaptive and ultrabroadband thermal control with solid-state nanophotonic emitters

## AUTHORS

Daniel Kindem[1*], Sam Keller[1*], Karl Pederson[1*], Yujie Luo[1], James Flaten[2], Ognjen Ilic[1]

[1] Department of Mechanical Engineering, University of Minnesota, Minneapolis, 55455 USA

[2] Department of Aerospace Engineering and Mechanics, University of Minnesota, Minneapolis, 55455 USA

## ABSTRACT

Managing the emission and absorption of thermal radiation is crucial for a wide range of technologies, from radiative cooling of buildings and vehicles to thermal regulation of satellites and future lunar and Mars habitats. Despite this universal and critical need, thermal emitters capable of adaptively modulating emissivity in a broadband, high-contrast, and fully solid-state manner remain elusive. Here, we leverage neural-network-guided photonic design to enable adaptive, solid-state thermal emitters based on chalcogenide phase-change materials capable of emissivity switching with extreme spectral contrast and bandwidth. These engineered nanophotonic emitters operate over a broad spectrum—from solar through thermal infrared—providing very low solar absorptivity while enabling switchable thermal infrared emissivity with high contrast. We experimentally demonstrate the core functionality of our approach in the space-like radiative environment in the stratosphere, observing a 31.5 °C temperature differential between the two solid-state phases of a simplified chalcogenide GeSbTe-225 thermal emitter. Our results point to even more significant capabilities, such as the potential to modulate >600 W/m$^2$ of radiative heat (at 100°C) with minimal solar heating in the vacuum of space. The proposed nanophotonic solid-state adaptive emitter could provide high-power and high-speed heat modulation while requiring no power to maintain state, offering transformative capabilities for thermal control in dynamic radiative environments on Earth and in space.

Radiative thermal management relies on deliberately engineering the balance of absorbed and emitted light across the electromagnetic spectrum—from ultraviolet through infrared—with applications spanning radiative cooling on Earth to heat transfer in space[1]. This approach fundamentally depends on developing materials with spectrally selective optical properties. For example, various paints incorporating metal-oxide additives (such as zinc oxide or titanium oxide[2,3]) exhibit low absorptivity (high reflectivity) for solar wavelengths and high emissivity for infrared wavelengths, acting as radiative coolers by reflecting sunlight while emitting thermal radiation[4]. More recently, advances in nanophotonics and nanofabrication have enabled engineered metamaterials[5–7] and metasurfaces[8,9] capable of enhancing spectral selectivity through precise control over the size and arrangement of nanoscale features[10–12]. So far, the majority of spectrally-selective structures are static: once fabricated, their emissivity state cannot be changed, preventing them from adapting to a dynamic thermal environment. By contrast, the search for high-performance variable emitters, capable of high contrast between emissivity states, has pursued several approaches. Mechanical approaches to variable thermal emission include both small-scale louvers and large-scale deployable radiators[13–15]. These systems can effectively manage heating by changing between reflective and absorptive configurations[16,17]; however, they require mechanical actuators, leading to more weight, slow response, and higher chance of mechanical failure. Alternatives such as electrochromic materials can change optical properties depending on applied voltage[18] and materials such as tungsten trioxide ($WO_3$) have been studied for infrared emissivity modulation[19,20], but the drawbacks include undesirable infrared absorption bands that limit modulation contrast, temperature sensitivity of ionic intercalation, and the need for constant power for electric field-driven modulation[21]. On the other hand, thermochromic materials, such as vanadium dioxide ($VO_2$), can transition between metallic and insulating states in response to a change in temperature[22–24]. This transition in $VO_2$ was harnessed for the design of variable emitters operating around the phase transition temperature (~67 °C)[25–28]. However, many applications require variable emissivity at different operating temperatures (or over a range of temperatures), but tuning the $VO_2$'s transition temperature—usually achieved by doping—significantly diminishes the emissivity contrast between metallic and insulating states[29]. In Supplementary Materials, we include a performance comparison of representative thermochromic works from the literature (Figure S1 and Table S2).

Here, by combining multi-state, multi-objective neural-network guided photonic design with chalcogenide phase-change materials, we present a variable thermal emitter capable of emissivity switching with extreme contrast and bandwidth. As a concept proof of this approach, we focus on variable emissivity in a broadband, space-like environment, predicting structures that simultaneously suppress solar heating and strongly modulate thermal-infrared emission between phase-change material states. Experimentally, we develop a stratospheric test platform that reproduces key aspects of the space radiative environment and use it to validate the emissivity behavior of simplified fabricated samples. During a test flight in this unattenuated radiative environment above the bulk of the atmosphere (**Figure 1**), we measure a large temperature contrast between amorphous and crystalline phases of the nanoscale chalcogenide layer.

We focus on chalcogenide alloys for the active layer in the metamaterial heterostructure as they exhibit an externally switchable phase-change between amorphous and crystalline states accompanied by a large refractive index modulation. The rich compositional space of Ge-Sb-Te alloys has been harnessed in memory and optical storage applications[30], photonic circuits[31,32], active optics[33,34], and spectroscopic sensing[35], sparking growing interest in related chalcogenide compositions such as GeSbSeTe[36] and SbSe[37]. The large modulation of infrared optical properties is particularly useful for emissivity and thermal camouflage applications[38,39], which further benefit from the ultrafast phase-change switching speeds, stability over a broad temperature range, and ability to maintain state without requiring a continuous supply of power. We note that chalcogenide materials have proven capable of switching state over a million cycles[40]. Recently, low Earth orbit tests through the Materials International Space Station Experiment-14 (MISSE-14) validated chalcogenide alloys' resistance to radiation, highlighting their potential for space-borne applications[41].

**Figure 1** highlights the dominant radiative heat channels from solar, albedo, and Earth radiation. Heating from the Sun is both strong and highly variable; for example, a satellite in low Earth orbit can enter and exit an eclipse more than fifteen times a day (approximately every 1.5 hours). The spectral response of an optimal emitter, illustrated in **Figure 1b**, requires multi-band functions: it suppresses the impact of solar heating by having a high reflectivity (low emissivity) for solar wavelengths; conversely, it actively modulates output radiation by switching between thermal-infrared emissivity states with high contrast. Such an emitter in a fully solid-state form (i.e., with no moving parts) could have a transformative impact, but there has yet to emerge a high-performing design that meets these requirements.

## DESIGN OF THE NANOSTRUCTURED EMITTER

We develop an adaptive metamaterial emitter by combining a chalcogenide GeSbTe phase-change layer with a neural-network inverse photonic design approach. Our emitter can mimic the ideal profile of **Figure 1b**, achieving highly broadband, spectrally-selective, and spectrally-switchable properties with just a single active chalcogenide layer. Highlighted in **Figure 2**, the emitter design comprises a chalcogenide phase-change layer placed on top of a metallic substrate. The metallic layer serves two functions: it ensures the emitter maintains broadband opacity while also enabling state switching by acting as an electrode. Above the phase-change layer, we construct a photonic environment using multiple dielectric layers selected from a predefined material set. **Figure 2a** depicts our neural network framework for determining optimal material arrangement and layer dimensions, including the phase-change layer, aimed at minimizing solar absorptance ($\alpha_{sol}$) and maximizing the thermal emissivity contrast ($\Delta\varepsilon_{th}$) between chalcogenide states. Here, the solar absorptance is $\alpha_{sol} = \int \alpha_\lambda I_{b\lambda}(T_{sol})d\lambda / \int I_{b\lambda}(T_{sol})d\lambda$, and the emissivity contrast is $\Delta\varepsilon_{th} = \varepsilon_{th,2} - \varepsilon_{th,1}$ where $\varepsilon_{th,1(2)}$ refers to the emissivity in the state "1" ("2") of the chalcogenide layer and $\varepsilon_{th} = \int \varepsilon_\lambda I_{b\lambda}(T_{surface})d\lambda / \int I_{b\lambda}(T_{surface})d\lambda$ is the hemispherical-total emissivity. Neural networks, such as Generative Adversarial Networks (GANs), are attracting significant interest for designing and optimizing optical components and photonic devices including for designing thin-film multilayer structures[42]. We extend this approach and make it compatible with the situation where one of the layers can exist in two states, crystalline and amorphous, with different complex refractive indexes. Conceptually, we train two parallel networks, one optimized for low solar absorptance and another for strong emissivity contrast, combining and subsequently refining their output into a single multi-band variable emitter structure (**Figure 2a**). We optimize computation time by calculating the complex intermediate impedance corresponding to the response of the static layers—which represents the majority of the calculation—only once, then reuse it for both amorphous and crystalline state calculations. The design approach is presented in the Methods section.

**Figure 2b** shows the spectrum of a variable metamaterial emitter designed using this approach. The emissivity spectrum shows very high hemispherical reflectance for solar wavelengths (≈0.93) and a very high emissivity contrast for thermal-IR wavelengths: a $\Delta\varepsilon_{th} \approx 0.80$ absolute emissivity difference between high/low emissivity and ≈12:1 ratio of high-state emissivity to low-state emissivity over the shown spectral range, closely mimicking the desired ideal performance of **Figure 1b**. The plotted spectrum is hemispherical (integrated over all angles). As a proof of concept of the proposed design approach, the emitter was designed using two neural networks (trained on 10 and 80 layers, respectively) for a 71-layer stack of GeSbTe-225, Ge, $BaF_2$, and ZnSe (see Methods and Supplementary Table S1 for the material and layer arrangement). For context, Supplementary Figure S1 and Table S2 present a comparison of these results with representative works from the literature. Importantly, the optimized emitter demonstrates high performance under varying environmental conditions. For the case of a planar emitting surface in low Earth orbit subject to solar irradiation, **Figure 2c** shows the effect of the incident solar angle on the dissipated heat. First, under direct exposure to sunlight, the emitter maintains consistent heat dissipation performance in both crystalline and amorphous states, as indicated by the solid blue and red lines in Figure 2c. Second, in terms of heat rejection (i.e., the balance between emitted and absorbed radiation), the crystalline state of the emitter consistently and significantly outperforms a blackbody emitter over a broad range of incident sunlight angles. The crystalline state's heat dissipation flux exceeds that of a blackbody emitter by more than 1100 $W/m^2$ at normal incidence, maintaining superior dissipative properties even at oblique angles of solar incidence up to 82 degrees. Similarly, the amorphous state consistently maintains the desired properties of low dissipation and high solar reflectivity (low solar absorptivity).

**Figure 2d** illustrates how integrating the adaptive emitter in an orbiter can effectively compensate for varying heat flux while in a low Earth orbit. As an example, we model a (spherical) orbiter generating 150 W of power to emulate a high-power small satellite. We analyze the performance of the metamaterial emitter compared to a blackbody-like surface. The blackbody surface absorbs excessive solar irradiation when facing the sun while overcooling when in eclipse. In contrast, the adaptive metamaterial emitter maintains the desired high heat rejection (in crystalline state) and low heat rejection (in amorphous state) throughout the orbit. Combining these features, we show how an actively controlled response can maintain a near-uniform temperature throughout the orbit (**Figure 2d**, purple). To achieve this active response, we assume a grid-based switching architecture, which is detailed in the Supplementary Information. This example demonstrates the potential of the proposed metamaterial emitter with highly contrasting emissive states for active thermal management in a dynamic radiative environment that changes between intense sunlight and eclipse.

## NANOFABRICATION AND FLIGHT DEMONSTRATION

**Figure 3** shows the results of fabrication, characterization, and theoretical prediction of a simplified proof-of-concept emitter. While the inverse-designed emitter in Figure 2 is presented as a predicted high-performance design, our first flight experiments focus on a single GeSbTe-225 layer to isolate and validate its state-dependent radiative behavior in a space-like environment (**Figure 4**). The thickness of the layer is designed to maximize the temperature contrast between the crystalline and amorphous phases (see Methods). **Figure 3b** shows the measured FTIR spectra of the structure for different annealing temperatures. Combining UV-Vis and FTIR spectroscopy, we characterize the emissivity of the structure from the visible through the infrared spectrum, observing good agreement with the emissivity predicted by the thin-film model (**Figure 3c**). As desired, we observe a strong contrast in the thermal-infrared emissivity between the crystalline and amorphous states, with the crystalline state as the high heat dissipation state. Additionally, the crystalline state is more reflective (less emissive) for solar wavelengths, resulting in less sunlight absorption and, overall, a lower equilibrium temperature compared to the amorphous state. **Figure 3d** shows the fabricated emitter samples integrated into the payload experiment. The payload experiment setup consists of a row of identical emitters prepared in crystalline and amorphous phases, respectively, suspended over a frame. A camera is positioned adjacent to the emitters and oriented in the same direction to record the radiative environment seen by each sample. An identical set of emitters and a second camera are installed on the opposite side of the payload, ensuring reliable data collection in the case of instrument malfunction. This redundant configuration reduces the risk of data loss during the inflight experiment.

**Figure 4** shows the configuration and the results of our flight experiment. In a flight over the state of Minnesota, USA, the payload stack reached an altitude of ~101,000 feet (31 km). The stack, displayed in **Figure 4a**, comprised of a venting unit (labeled "A"), a 360-camera unit (labeled "B"), and the experiment payload (labeled "C", see also inset). The 360-degree background photo taken by the camera unit is indicative of the fact that the space-Sun-Earth radiative environment at these altitudes resembles that of a low Earth orbit, unattenuated by the atmosphere. **Figures 4b,d** illustrate the custom onboard electronics system for data acquisition. At its core is a Teensy microcontroller that interfaces with several sensors: a high-altitude-enabled GPS, an inertial measurement unit (IMU), a barometric altimeter, and an XBee radio module for short-range intra-stack communication (see Methods for details). Temperature measurements of emitter samples are collected by a set of thermistors. In addition to interfacing with the listed sensors, the microcontroller broadcasts and logs a 4-bit signal to the four LEDs placed in the field of view of the camera (lower right corner of camera insets in **Figure 4e**). We broadcast a predetermined, non-repeating sequence refreshed at 1 Hz rate to act as a visual timestamp and use this information to synchronize the microcontroller timeline with the camera timeline. This approach allows us to accurately match each camera frame with its corresponding sensor readings from the microcontroller.

We determine the optimal launch time and orientation of the emitter surface by analyzing the incident direction of sunlight on the emitters in flight. **Figure 4c** shows the cosine of the incident sunlight angle as a function of the sample emitter tilt angle ($\beta_S$) and the horizontal angle of the emitter's projected surface normal ($Z_S$). Angle $Z_S$ corresponds to the rotation of the payload stack about the vertical axis (gravity axis): it ranges from 0 to 360 degrees and varies during flight. A description of the orientation and the governing coordinate system is presented in the Supplementary Materials (**Figure S2**). For the planned location and date of flight, the optimal emitter tilt angle $\beta_S \sim 25°$ was determined at peak altitude. The payload box was constructed to tilt the emitter plane at this angle, ensuring direct sunlight hits the emitter plane when the payload reaches the target altitude. The camera mounted pointing normal to the emitter plane validates this construction by showing the sun in the center of the frame.

**Figure 4f** shows the experimental in-flight temperature measurements of each emitter, demonstrating a consistent and substantial thermal difference between the emitter's amorphous and crystalline states regardless of the payload's rotation relative to the Sun. We observe a peak temperature difference of 31.5 °C between the GST solid-state phases during increased solar exposure (**Figure 3c**), consistent with the transient lumped-capacitance heat transfer model presented in the Methods and Supplementary Materials. Starting from approximately 90,000 ft altitude, a vent is periodically opened to release lift gas and reduce both the ascent rate and the rotational motion of the stack. Finally, after near-float around the altitude of ~100,000 ft, the balloon is intentionally detached from the payload, initiating rapid descent (**Figure 4f**). The bottom two frames, shown in **Figure 4e**, provide examples of the radiative environment when samples face away from and toward the Sun and the corresponding temperatures in those instances. By matching individual camera frames with temperature measurements, we confirm that sample heating directly correlates with solar orientation.

## CONCLUSION

These results suggest that variable emitters based on solid-state phase switching may provide an effective platform for thermal management, including thermal management in space. Through experiments conducted in the radiative environment of near space, we found that even simple single-layer emitters achieved significant thermal contrast, achieving a 31.5 °C temperature difference between the emitter phases. Similarly, our design predicts the potential for more significant capabilities, including thermal dissipation rates of 800 W/m$^2$ in direct sunlight at 100 °C. To facilitate our measurements, we developed a novel experimental platform that allows us to access the radiative environment of space at a fraction of the cost and with the potential for more rapid iteration compared to traditional CubeSat deployments in low Earth orbit.

We envision several relevant extensions of this work. On the experimental side, the incorporation of a faster response venting mechanism, or a fully autonomous venting mechanism, would enable longer flights and extend measurement time. In addition, adding a ballast system could allow for active control of altitude. For rotation control, the use of reaction wheels or geomagnetic alignment could allow precision pointing and the ability to track or avoid sunlight as needed for testing. We further envision the incorporation of electronic hardware to allow for intentional switching of emissivity in flight. In an adaptive implementation, the metallic backplane can serve as a bottom electrode, switching PCM state via locally induced Joule heating. The intentional use of only a single PCM layer and its direct placement on the metallic backplane is beneficial for switching and helps to mitigate thermal effects from the rest of the dielectric stack. Reported switching times can be ultrafast (e.g., μs-scale[41]) with energy estimates reported similar to electrochromic devices (e.g., ~7 J for a 300-cm$^2$ panel[43]). Furthermore, exploring and demonstrating a grid-based switching architecture[44] that allows sequential switching of different sections could be a mechanism for continuous, fine-tuned control of emissivity in flight.

Our design and flight experiments focused on one chalcogenide alloy (GeSbTe-225), but the growing family of similar solid-state phase-change materials, including GSST, SbSe, GSTT, presents a promising material design space for space radiative applications that remains to be fully explored. Here, our design approach from **Figure 2** can be expanded to treat the PCM layer material as an optimization parameter, beginning with a continuous probability distribution across a list of PCM candidates before the neural network settles on the discrete (single) optimal PCM choice. This framework would enable optimization of the PCM material selection but could also be used to inform how intermediate states (states between crystalline and amorphous in a given PCM) might be optimally used in the thermal design. Furthermore, it would be important to understand how to balance emitter performance and strong emissivity contrast against structural complexity and number of layers, potentially revealing simpler yet effective emitter designs. The ability to control radiative thermal properties across a broad spectrum spanning from solar to thermal-infrared wavelengths, with high fidelity and adaptability, is crucial for solving the thermal challenges we face on Earth as well as in space – from enabling smaller and more powerful CubeSats to supporting our aspirations for sustainable habitats on the Moon and Mars.

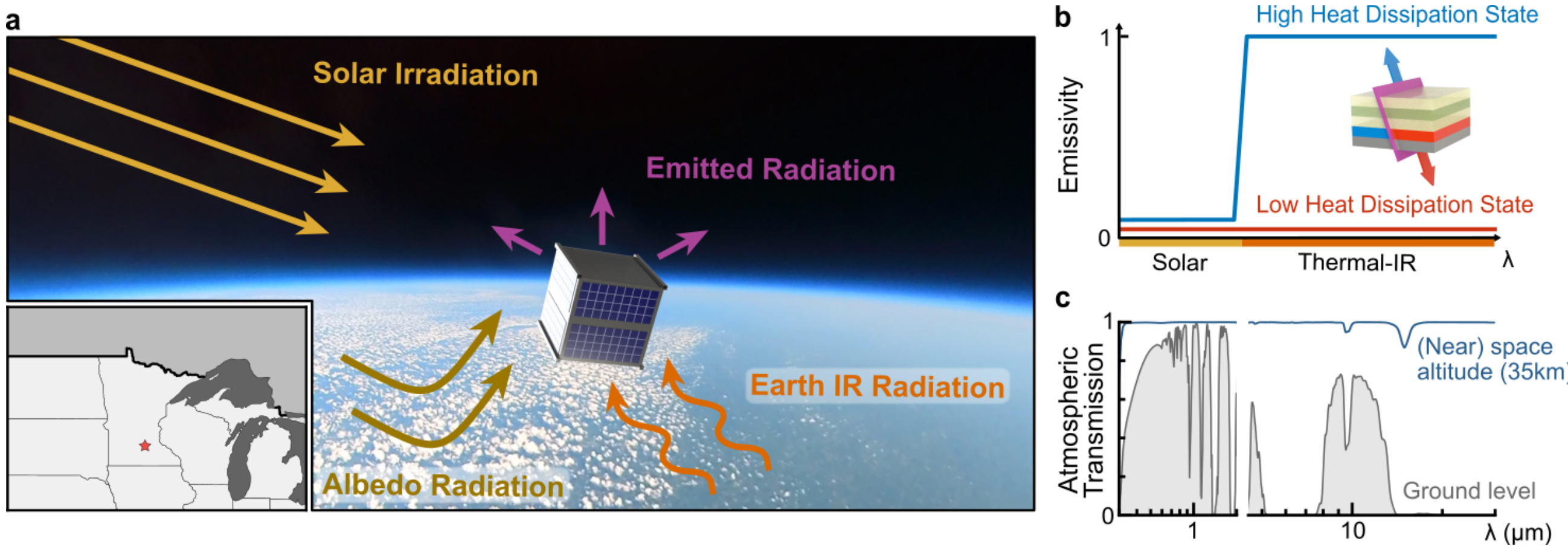

**Figure 1: a.** Radiative heat loads on a concept object. The background image was taken on a near-space stratospheric flight where the experiment was conducted (see Video 1 and Figure 4). Inset: launch location of the flight. **b.** Ideal adaptive radiator will have high solar reflectance (low solar emissivity) coupled with large emissivity modulation between the high- and low-heat dissipation states in the thermal-infrared spectrum (Inset: metamaterial radiator with an active solid-state phase-change layer). **c.** Comparison of atmospheric transmission at (near) space altitude and ground level, indicating that the radiative environment at stratospheric altitudes resembles that of outer space, as also evident from the background image in panel (a) and Video 1.

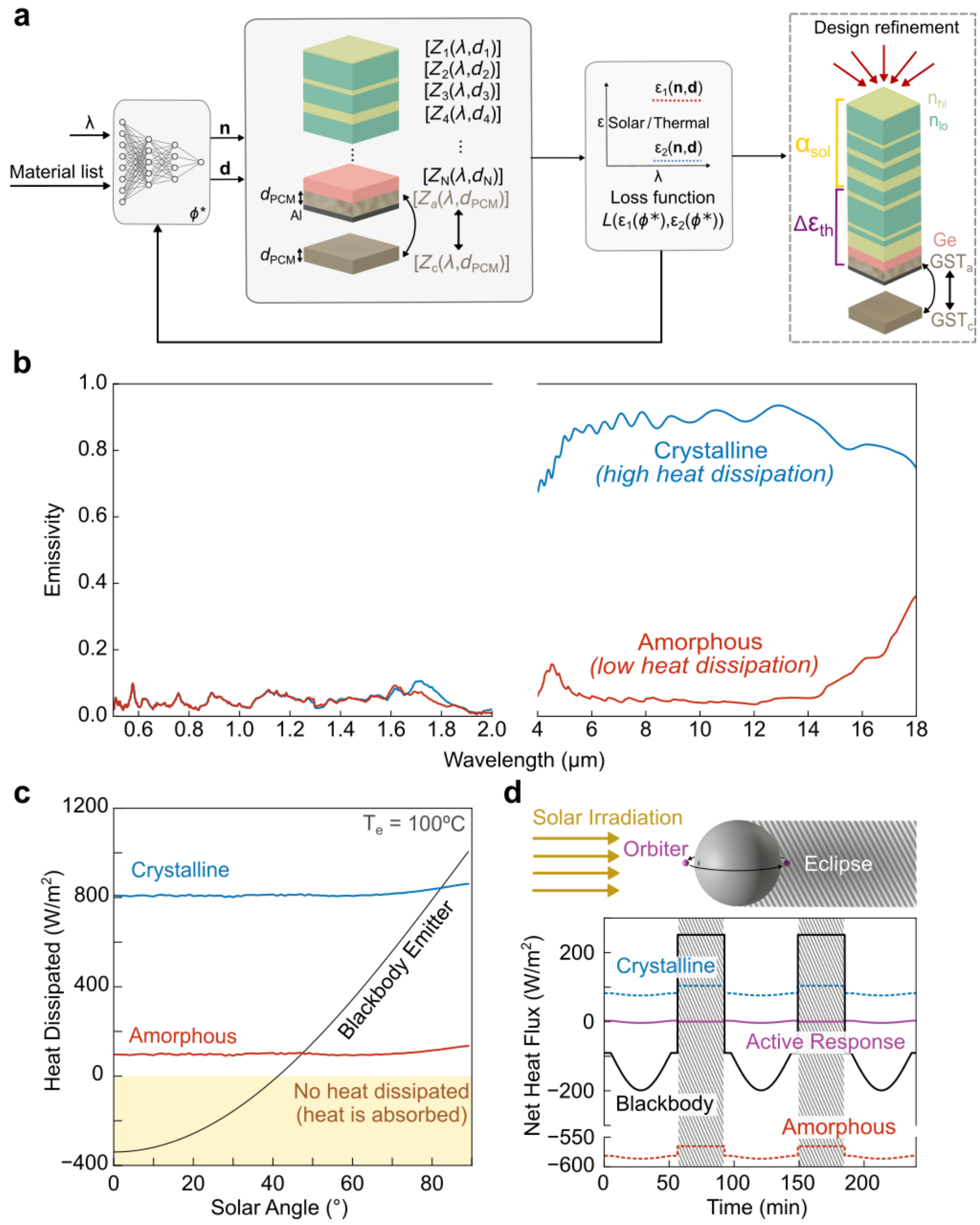

**Figure 2: a.** Neural-network driven photonic design of the two-state adaptive emitter. **b.** Spectral optical properties of the designed metamaterial emitter resemble the target response of the ideal adaptive emitter from Fig. 1b. Plotted emissivity is hemispherically averaged. For context, Supplementary Figure S1 and Table S2 present a comparison of these results with representative works from the literature **c.** Comparison of heat dissipation capabilities of the emitter in its two solid states, crystalline (blue) vs amorphous (red) as a function of the solar angle. A blackbody emitter is included for comparison. **d.** Heat flux analysis for an example low Earth orbit, in sunlight vs eclipse. Combining high- and low-heat dissipation states, an actively controlled response can maintain a net zero heat flux throughout the orbit (purple).

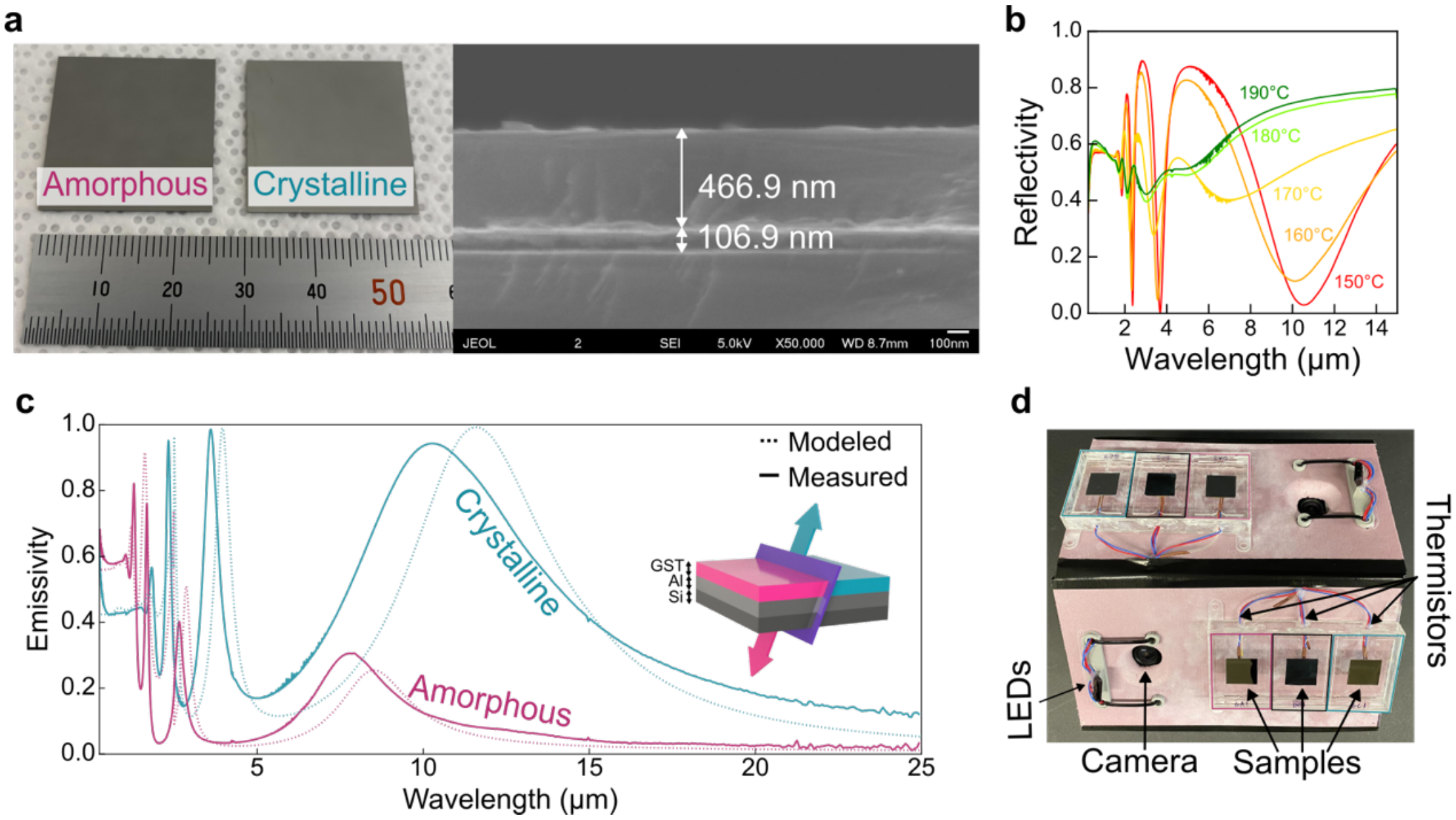

**Figure 3: a.** Fabricated active phase-change layer in amorphous and crystalline state (GST-225 on top of a thin aluminum layer). **b.** Reflectance change as a function of annealing temperature showing the phase-change transition. **c.** Measured and predicted emissivity of the active phase-change layer in its two states. **d.** Assembled experimental payload with outside-facing samples for dynamic temperature measurement.

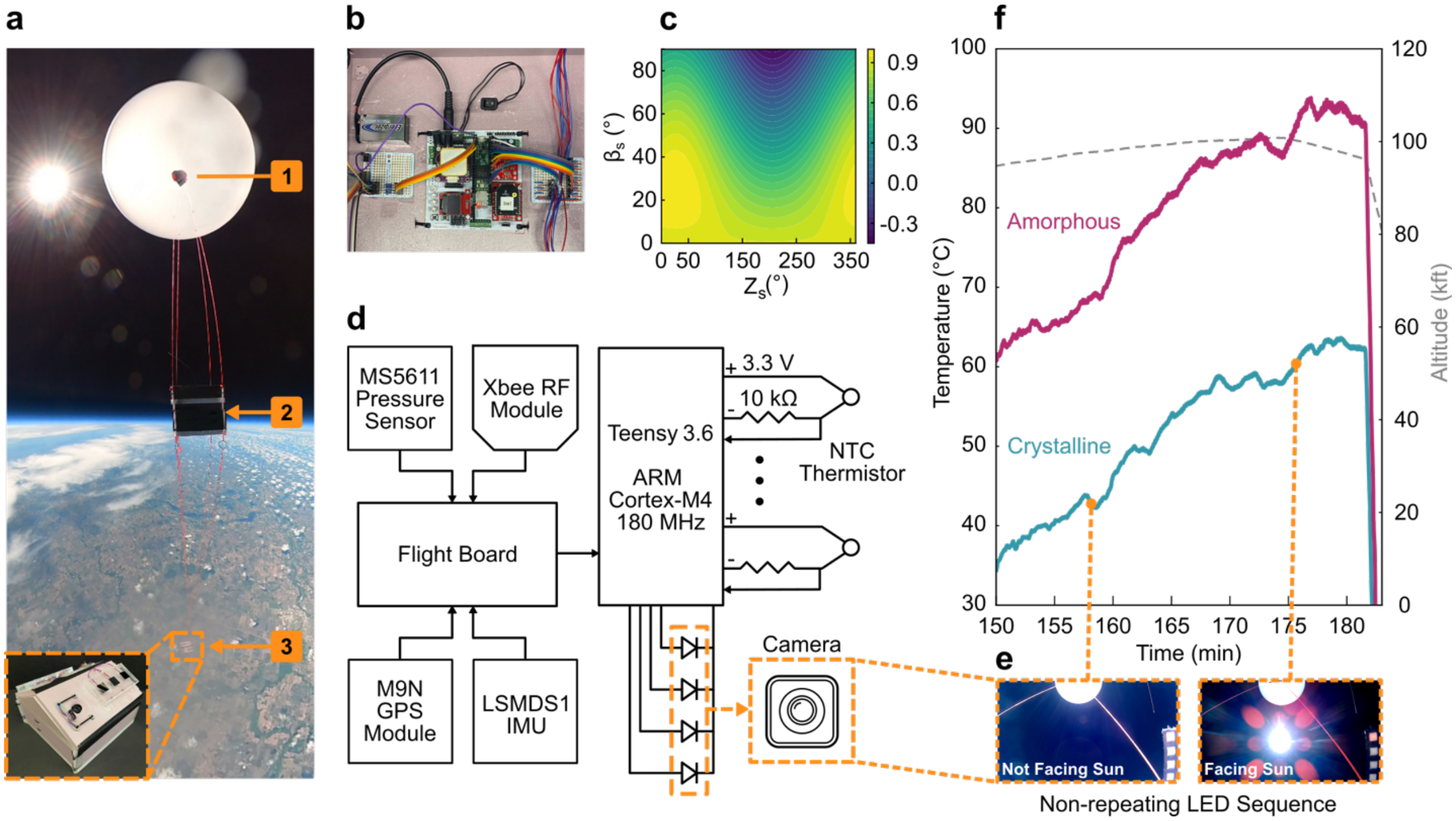

**Figure 4: a.** Flight results for proof-of-concept samples. Stratospheric test platform consisting of a (1) balloon and venting system, (2) imaging payload, and (3) thermal measurement payload. **b.** Flight computer and data acquisition electronics contained within the thermal measurement payload. **c.** Solar intensity as a function of the sample azimuthal angle ($Z_s$) and the sample tilt angle ($\beta_s$) (refer to Supplementary Figure 1). **d.** Electronics schematic of the thermal measurement payload where LEDs are used for image-data synchronization. **e.** Camera frames highlighting the radiative environment facing towards and away from the sun as the suspended payload slowly rotates. **f.** Temperatures of crystalline and amorphous samples measured at peak altitude.

*These authors contributed equally.
E-mail: ilic@umn.edu